\documentclass[aps,pre,twocolumn]{revtex4-2}
\usepackage[utf8]{inputenc}
\usepackage{bm}
\usepackage{graphicx}
\usepackage{amssymb}
\usepackage{mathtools}
\usepackage{amsmath}
\usepackage{xcolor}
\usepackage{enumitem}
\usepackage{natbib}
\usepackage{hyperref}
\usepackage{lipsum}
\usepackage{appendix}
\usepackage{color}

\begin{document}
\title{Network Reconstruction in Consensus Algorithms with Hidden Agents}
\author{Melvyn Tyloo}
\affiliation{Living Systems Institute
and} 
\affiliation{Department of Mathematics and Statistics, Faculty of Environment, Science, and Economy, University of Exeter,
Exeter, United Kingdom}
\date{\today}
\begin{abstract}
{Reconstructing the parameters that encode the influence between model variables based on time-series measurements represents an outstanding question in the theory of complex network-coupled systems. Here, we propose a solution to this problem for a class of noisy leader-follower consensus algorithms, where one has access to measurements only from the followers but not from the leaders. Leveraging the directed Laplacian coupling of such systems, we present an autoregressive expansion of the observed dynamics which can be truncated at different orders, depending on the memory of the leaders.
When their memory is short, this allows one to correctly reconstruct the full dynamical matrix with hidden leader agents, provided some additional assumption on the system to lift the degeneracy in the reconstruction. 
We illustrate and check the theory using numerical simulations for the cases of both a single and multiple hidden leaders. }
\end{abstract}

\maketitle

\section{Introduction}
Networked systems find numerous physical and engineered realizations such as large-scale power transmission networks, chemical reactions, proteins folding or even social interactions on influence networks and autonomous vehicles flocking. 
They are made of individual dynamical systems with their own internal parameters, which somehow interact together~\cite{Str04,Bar16,New18book}. 
The time-evolution of such systems is essentially dictated by the overall interplay between internal dynamics, coupling structure and external influence from the environment~\cite{Pik03}. 
Due to their high dimension, such systems are usually impossible to fully monitor because of cost constraints or simply because some elements are not accessible~\cite{Wan16b,Bru18,Bra03,succar2025detecting}. 
However, in order to control a networked system and prevent potential failures, it is highly desirable to know its parameters including both monitored and unmeasured elements. 

The inference of model parameters from time-series measurements represents an outstanding problem in network theory\cite{ljung1998system,New18}. 
Even more challenging is the case where not all agents are monitored. Recent methods based on optimization of local likelihood functions allow to infer the covariance between agents~\cite{Dan19}. 
For diffusively coupled systems, various situations where time-series are obtained from multiple initial conditions~\cite{Tim07} and from a system subjected to ambient noise~\cite{Ren10,zhang2015solving,Tyl21a,vu2025symmetric} or probing signals~\cite{Tyl21,Del21} have been considered when all the agents are monitored.
 One is typically able to reconstruct the full connectivity by pseudo-inversion of the correlation matrix~\cite{Ren10}. 
 In the more complex case where only a subset of agents are monitored, it is still possible to reconstruct the connectivity within this subset leveraging the correlation matrix of time derivatives of the degrees of freedom~\cite{Tyl21a}. 
 In general, without additional information, it is not possible to infer the full network connectivity solely based on time-series coming from monitored agents, mostly because of degeneracies in the reconstructed matrices. 
 However, in various realistic settings, we may have information on the agents that are not measured. For example, one may know that the hidden agents are only a small fraction of the total nodes, not interacting with one another. Or one may know the overall structure of the networked system, but do not have measurements everywhere in the system. The additional information included in the latter cases may enable a full reconstruction of the coupling matrix. 

In this Letter, we focus on a noisy linear leader-follower consensus algorithm where the coupling is given by a directed Laplacian matrix.  
Such dynamics is similar to multi-agent consensus algorithms ~\cite{saber2003consensus,hong2008distributed,Pat10} and opinion formation models~\cite{taylor1968towards,baumann2020laplacian}\,.
Assuming access to measurement time-series only from the followers, we leverage an autoregressive expansion of the observed dynamics to infer a collection of matrices that are given by products of the blocks of the overall dynamical matrix of the system. 
 {Importantly, Ref.~\cite{nozari2017network} showed that one can reconstruct the terms of such autoregressive expansion and thus also infer the direct connections between the measured nodes. 
Here, we want to go one step further and recover not only the coupling among the observed nodes, but the entire dynamics.}
For a single hidden leader, we show that one can fully reconstruct the dynamics when the leader has a short memory. 
When there are more than one hidden leaders, one needs to add additional assumptions to circumvent the degeneracy of the reconstructed dynamics. 
Namely, to achieve that, we require that the leaders are not connected to the same observed follower, that their coupling to the followers is symmetric. 
{The present work goes beyond the previous literature as it does not assume full observability or homogeneous noise, and under specific assumptions, enables the full reconstruction of the original leader-follower dynamics.}

\section{Consensus algorithm}
As dynamical system, we consider a network of $N=N_f + N_l$ agents with degrees of freedom $x_i\in\mathbb{R}$ for $i=1,...N$, interacting via a discrete leader-follower consensus dynamics. 
{We assume that there are more followers than leaders, i.e. $N_f>N_l$\,.}
Their time-evolution is described by the following coupled maps,
\begin{align}\label{eq1a}
{{x}_i}(t+\Delta t) &= x_i(t)- \sum_{j=1}^{N}\,k_{ij}\,[x_i(t) - {x_j}(t)] + \xi_i(t) \,,\,
\end{align}
for $i=1,...N_f$\,, and,
\begin{align}\label{eq1b}{{x}_i}(t+\Delta t) &= \alpha_i x_i(t) - \sum_{j=1}^{N}\,k_{ij}\,[x_i(t) - {x_j}(t)] \,,\, 
\end{align}
for $i=N_f+1,...N$\,, 
where Eq.~(\ref{eq1a}) corresponds to the observed followers and Eq.~(\ref{eq1b}) to the hidden leaders. 
Without loss of generality, we assume a vanishing initial condition i.e. $x_i(0)=0$ for $i=1,...N$\,. 
The coupling among the agents is given by the elements $k_{ij}\ge 0$ of the adjacency matrix $K\in\mathbb{R}^{N\times N}$ which is not assumed to be symmetric. 
The last term in the follower dynamics $\xi$ represents Gaussian white-noise inputs, i.e. $\langle \xi_i(t)\xi_j(t') \rangle = \xi_{i,0}^2\delta_{ij}\,\delta_{tt'}$\,, with $\xi_{i,0}$ not necessarily the same everywhere. 
Note that leaders are noiseless but have an additional term with $|\alpha_i|\le1$ that tends to bring their degree of freedom to zero. Note that when $\alpha_i=1$\,, the behavior of the $i$-th leader is the same as a follower. 
We assume that the overall system parameters are such that, for long enough times, the dynamics is fluctuating around the consensus state given by $x_i(t)=0$ for $i=1,...N$ in the noiseless deterministic case. 
It is insightful to rewrite the consensus dynamics in a matrix form as,
\begin{align}\label{eq2}
{\bf x}(t+\Delta t) = 
\begin{bmatrix}
{{\bf x}_o}(t+\Delta t)\\
{{\bf x}_h}(t+\Delta t)
\end{bmatrix}
=
\underbrace{\begin{bmatrix}
B & C\\
D & E 
\end{bmatrix}}_{\bf{A}}
\begin{bmatrix}
{{\bf x}_o}( t)\\
{{\bf x}_h}( t)
\end{bmatrix}
+
\begin{bmatrix}
{\bm \xi}_o(t)\\
\bf 0
\end{bmatrix}.
\end{align}
 {The matrix elements of $B$ contain information about the coupling among the followers and their overall degree, i.e. $B_{ij} = \delta_{ij}(1 - \sum_{l=1}^N k_{il}) + k_{ij} $\,, $i,j=1,...,N_f$\,. Similarly, $E$ includes the information about the coupling among the leaders and their internal parameters, i.e. $E_{ij} = \delta_{ij} (\alpha_i) - \sum_{l=1}^{N_l}k_{i+N_f l+N_f} + k_{i+N_f j+N_f}$\,, $i,j = 1,...,N_l$\,. The remaining blocks $C$ and $D$ contain respectively the leader-to-follower and follower-to-leader coupling, i.e. $C_{ij} = k_{i j+N_f}$\,, $i=1,...,N_f$\,, $j=1,...,N_l$ and $D_{ij} = k_{i+N_f j}$\,, $i=1,...,N_l$\,, $j=1,...,N_f$\,. In the following, we focus on reconstructing these four matrices, keeping in mind that they are directly related to the coupling and internal parameters of the system.}
It is important to note that, because of the Laplacian dynamics Eqs.~(\ref{eq1a}), (\ref{eq1b})\,, the blocks of the matrix $A$ satisfy $\sum_{j=1\\ j\neq i }^{N_f} B_{ij} + \sum_{k=1}^{N_l} C_{ik} = - B_{ii}$ for $i=1,..., N_f$ and $\sum_{j=1}^{N_f} D_{ij} + \sum_{k=1}^{N_l} E_{ik} =  \Lambda_{ii}$ where $\Lambda_{ii} = \alpha_i$ for $i=1,..,N_l$\,. 
The latter conditions hold for rows but not for columns, as the dynamics we consider here is not necessarily symmetric. 
Eq.~(\ref{eq2}) gives the time evolution of the full state at $t+\Delta t$ as a function of the full state at time $t$\,. 
One can rewrite the time evolution of the observed part of the network using only the past of the observed agents. 

\section{Autoregressive expansion}
Iteratively expressing ${\bf x}_h(t)$ with ${\bf x}_o(t-\Delta t)$ and ${\bf x}_h(t-\Delta t)$\,, one can write the observed dynamics as,
\begin{align}\label{eq3}
    {\bf x}_o(t+\Delta t) &= B \,{\bf x}_o(t) + {\bm \xi}_o(t) + CD\, {\bf x}_o(t-\Delta t) \\
    + CED\, &{\bf x}_o(t-2\Delta t) + CE^2D\, {\bf x}_o(t-3\Delta t) + ...\nonumber\\
    = B \,{\bf x}_o(t) &+ {\bm \xi}_o(t) + \sum_{k=0}^M CE^kD\, {\bf x}_o(t-(k+1)\Delta t)\,,\nonumber
\end{align}
where $(M+1)\Delta t = t$\,. Such expansion is similar to a Mori-Zwanzig approach where the unobserved variables effectively enter the observed dynamics with a memory kernel~\cite{mori1965transport,zwanzig1973nonlinear,tyloo2024resilience}.
Eq.~(\ref{eq3}) conveniently expresses the dynamics of the observed agents at time $t$ in terms of all the states of the observed agents that have been visited since the initial condition of the system and is exact. 
But it assumes a considerable knowledge about the system, namely, the states of the observed agents since $t=0$ and the initial condition of both the observed and unobserved agents. 
One can relax these assumptions at the cost of considering more specific systems, which allow for an approximation of Eq.~(\ref{eq3}) by truncating its last term.
Indeed, one can take e.g. the Froebinius norm of the last term in Eq.~(\ref{eq3})\,, such that $\|\sum_{k=0}^M CE^kD\, {\bf x}_o(t-(k+1)\Delta t)\| \le \sum_{k=0}^M \|C\|\|E\|^k \|D\|\, \|{\bf x}_o(t-(k+1)\Delta t)\|$\,. 
As the spectral radius of $A$ satisfies $\rho(A)<1$ in Eq.~(\ref{eq2})\,, the terms $\|{\bf x}_o(t-(k+1)\Delta t)\|$ for $k=0,...M$ are finite. 
So, without any assumptions on $C$ and $D$\,, in order for a truncation of Eq.~(\ref{eq3}) at $k=m$ to be a valid approximation, one requires $\|E\|<1$ and sufficiently close to $0$ such that $\|E\|^m\cong 0$\,. 
If $\|E\|$ is sufficiently small, one can  one can approximate Eq.~(\ref{eq3}) keeping only the terms up to $k=1$\,, i.e.

\begin{align}\label{eqmain}
    {\bf x}_o(t+\Delta t) &\cong B \,{\bf x}_o(t) + {\bm \xi}_o(t) + CD\, {\bf x}_o(t-\Delta t)\\
    &+ CED\, {\bf x}_o(t-2\Delta t)\,\nonumber\,.
\end{align}
This expression does not require to have access to measurements starting at $t=0$\,, nor the knowledge of the initial conditions. 
The condition that $\|E\|^k\cong 0$ for $k>1$ typically holds when the leaders have a short memory, i.e. their trajectory does not depend too much on their previous state. 
If the leader agents do have a longer memory, one might consider additional terms in the sum of Eq.~(\ref{eq3})\,.
Now, let us see how one gets estimates for $B$\,, $CD$ and $CED$ from time-series measurements. 

\section{Matrix Reconstruction}
 {To reconstruct the matrices in Eq.~(\ref{eqmain})\,, we follow the approach of Ref.~\cite{nozari2017network}\,.} 
One has access to time-series measurements of the observed agents, i.e. the followers ${\bf x}_o(t)$ for $t=0,...,(N_t-1)\Delta t=T$\,. 
Extending the dimension of the state vector to include multiple time steps ${\bf X}(t+2\Delta t) = [{\bf x}_o(t+2\Delta t), {\bf x}_o(t+\Delta t), {\bf x}_o(t)]^\top$\,, 
one can rewrite Eq.~(\ref{eqmain}) as,
\begin{align}
    {\bf x}_o(t+\Delta t) \cong [B,\, CD,\, CED]\, {\bf X}(t) + {\bm \xi}_o(t)\,.
\end{align}
Then, by right-multiplying the latter equation by ${\bf X}^\top(t)$\,, and taking the average over the iterations, one obtains,
\begin{align}
    \Sigma_1 \cong [B,\, CD,\, CED]\, \Sigma_0 \,,
\end{align}
where we defined the matrices 
\begin{align}
    \Sigma_0 &= \frac{1}{N_t-3}\sum_{k=2}^{N_t-2} {\bf X}(k\Delta t){\bf X}^\top(k\Delta t)\\
    \Sigma_1 &= \frac{1}{N_t-3}\sum_{k=2}^{N_t-2} {\bf x}_o((k+1)\Delta t){\bf X}^\top(k\Delta t)\,.
\end{align}
The above expressions allow one to derive estimators for the matrices $B$,\, $CD$,\, $CED$ as,
\begin{align}\label{eqrc}
    [\widehat{B},\, \widehat{CD},\, \widehat{CED}]\, =  \Sigma_1 \Sigma_0^{-1}\,.
\end{align}
These estimators for the matrices can then be used to uncover the full network connectivity in the leader-follower dynamics. 
 {Ref.\cite{nozari2017network} showed that the error made by similar types of estimator as Eq.~(\ref{eqrc}) decreases when one includes more terms in the autoregressive expansion, provided that the time-series are long enough. }

In the following, we start by considering the simpler case where only a single leader agent is hidden. 
Then we move on to the more complex situation where multiple leaders are hidden and discuss how the actual network can be recovered.

\begin{figure}
    \centering
    \includegraphics[width=0.99\linewidth]{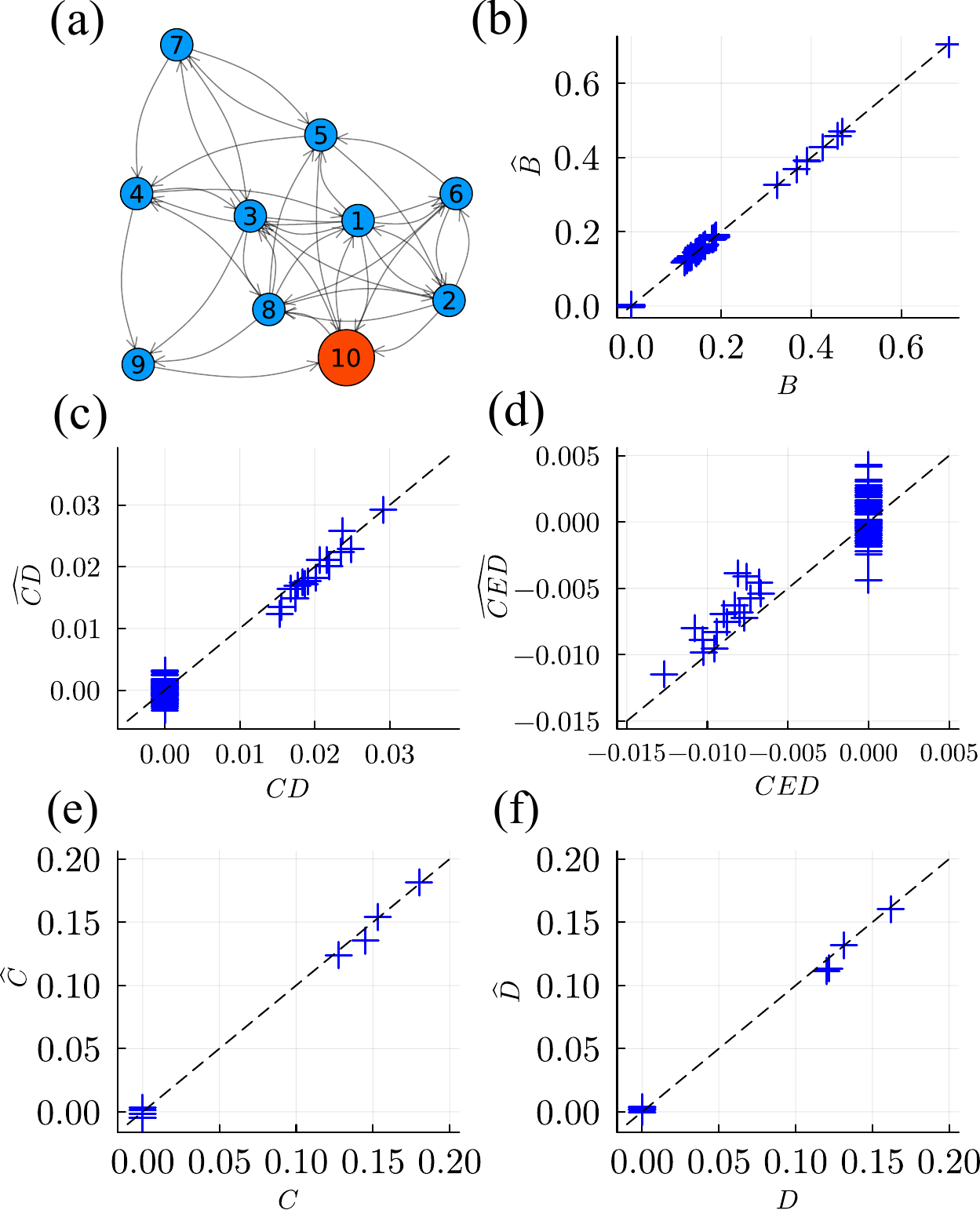}
    \caption{\textbf{Matrix reconstruction for a single hidden leader}. (a) Directed network of $10$ nodes with $N_f=9$ followers (blue) and $N_l=1$ hidden leader (red). 
    (b) Comparison between the actual coupling among the followers $B$ and the reconstructed one $\widehat{B}$\,. 
    (c) Comparison between the actual matrix $CD$ and the reconstructed one $\widehat{CD}$\,. 
    (d) Comparison between the actual matrix $CED$ and the reconstructed one $\widehat{CED}$\,. 
    The time-series used for the reconstructions were obtained by simulating the dynamics Eqs.~(\ref{eq1a}), (\ref{eq1b}) with the coupling given by the weighted network shown in panel (a). 
    The adjacency matrix of the network was obtained starting from a matrix of uniform random number between $0$ and $1$ and then keeping only the elements larger than $0.6$ and ignoring the diagonal. 
    The Laplacian matrix is then obtained from this adjacency matrix and normalized by the largest diagonal element. 
    The internal parameter of the leader is $\alpha_{10} = 0.1$\,, so that that $E=-0.435246$\,. 
    The length of the time-series is $N_t = 5\times 10^5$\,, with vanishing initial conditions. 
     {The noise standard deviation is 0.1\,.}}
    \label{fig1}
\end{figure}

\begin{figure}
    \centering
    \includegraphics[width=0.99\linewidth]{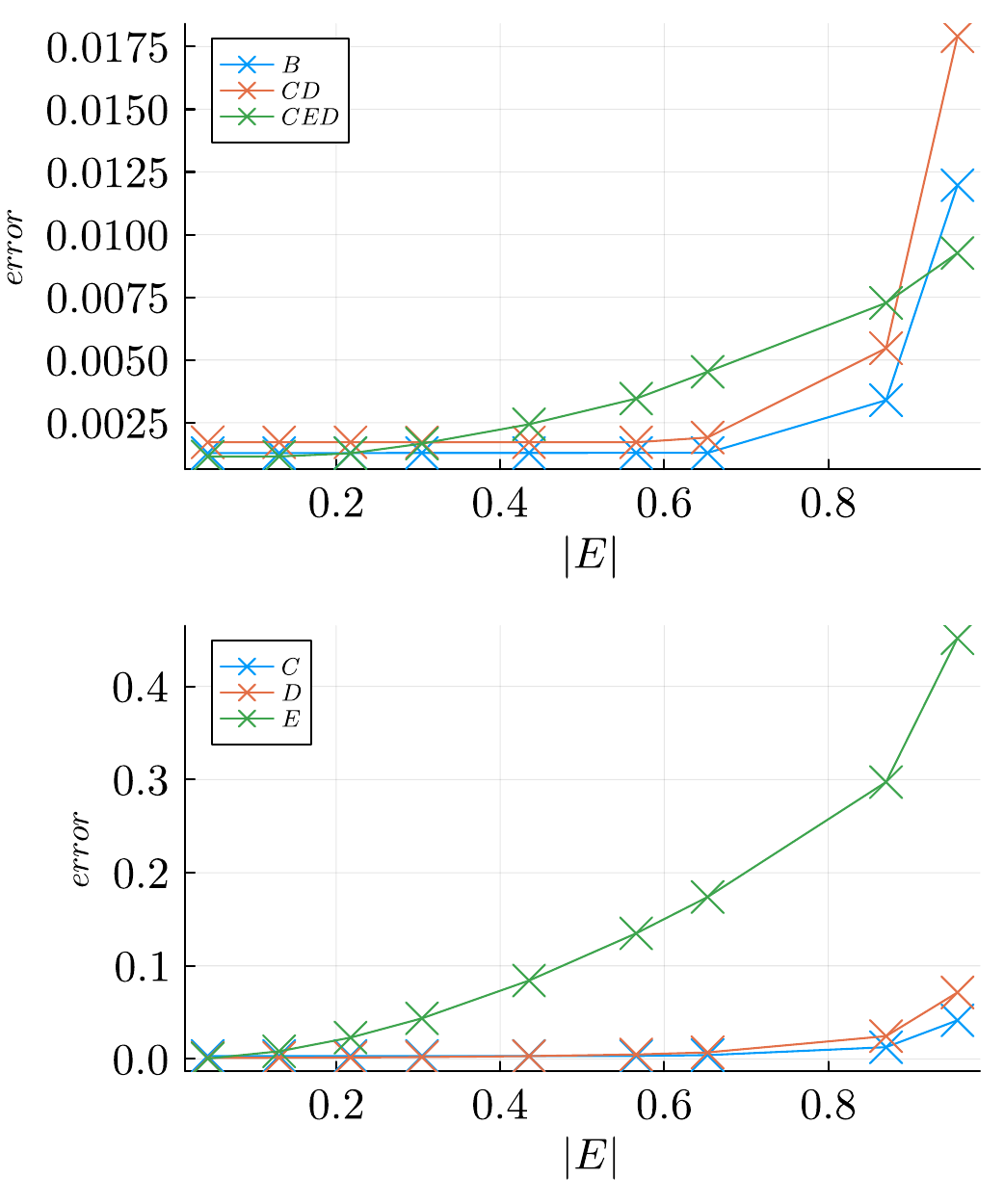}
    \caption{{\textbf{Error on matrix reconstruction for a single hidden leader}. Maximum of the absolute difference between the actual matrices and the reconstructed ones for different values of $E$\,. 
    The parameters of the system are the same as in Fig.~\ref{fig1}\,, but with $E$ that takes different values to test the validity of the approximation Eq.~(\ref{eqmain})\,. 
    The error can be compared to the absolute value of the non-vanishing elements of each matrix: $|B_{ij}|\in [0.118,0.707]$\,, $|[CD]_{ij}|\in [0.015,0.029]$\,, $|[CED]_{ij}|\in[0.0067,0.013]$\,, $|[C]_{ij}|\in[0.13,0.18]$\,, $|[D]_{ij}|\in [0.12, 0.16]$\,.    
    The length of the time-series is $N_t = 5\times10^6$\,, with vanishing initial conditions. }}
    \label{fig1_2}
\end{figure}
\section{Single unobserved leader}
Let us start with the easier scenario where there is only a single leader agent that is not observed in the system. 
Then, the blocks of $A$ are a $(N-1)\times (N-1)$ matrix $B$\,, a size $(N-1)$ column vector $C$\,, a size $(N-1)$ row vector $D$\,, and a scalar $E$\,. 
The condition so that Eq.~(\ref{eqmain}) is a valid approximation translates into $|E|=|\alpha_{N_l} - \kappa_{N_l}| \ll 1$ where we denoted the weighted in-degree of the hidden leader $\kappa_{N_l} = \sum_{j=1}^{N_f}k_{N j}$\,. 
It essentially depends on the internal drive of the leader back to the origin and its connectivity to the followers. 
The part of $A$ corresponding to the interaction within the followers is directly obtained by $\widehat{B}$\,. 
Leveraging the diffusive structure of the coupling in Eq.~(\ref{eq1a})\,, one can also obtain an estimate of the vector $C$ from $\widehat{B}$ as,
\begin{align}\label{est2}
    \widehat{C}_i = 1 - \sum_{j=1}^{N_f}\widehat{B}_{ij}\,,
\end{align}
for $i=1,...,N_f$\,. 
Having estimated $\widehat{C}$\,, one can reconstruct $D$ by solving the overdetermined system,
\begin{align}\label{est3}
    \widehat{C}\widehat{D} = \widehat{CD}\,,
\end{align}
to obtain $\widehat{D}$\,. 
Note that, in order to recover $\widehat{D}$ from the system Eq.~(\ref{est3})\,, one needs at least one non-vanishing component in $C$\,, which is implicitly assumed when deriving Eq.~(\ref{eq3})\,.
Because $E$ is simply a scalar here, one can get an estimate from the reconstructed matrices $\widehat{CD}$\,, $\widehat{CED}$ as,
\begin{align}\label{est1}
    \widehat{E} = M^{-1}\sum_{i,j\in\mathcal{M}}\frac{\widehat{CED}_{ij}}{\widehat{CD}_{ij}}\,,
\end{align}
where indices $i$, $j$ run over the set $\mathcal{M}$ of non-vanishing elements of $\widehat{CD}$\,, with $M=|\mathcal{M}|$ its cardinality. This set can be obtained be thresholding $\widehat{CD}$\,. 
Eventually, using $\widehat{D}$\,, one recovers the internal dynamics of the hidden leader $\alpha_{N_l}$ with
\begin{align}\label{est4}
    \widehat{\alpha_{N_l}} = \widehat{E} + \sum_{j=1}^{N_f} \widehat{D}_j
\end{align}
With Eqs.(\ref{est2})-(\ref{est4})\,, one can fully reconstruct the interaction network among the agent as well as the leader internal dynamics using only the measurements from the follower agents. 
Note that, after each of the reconstruction above, because the amount of data is finite, some matrix elements that are vanishing in the actual system might be inferred as non-zero, but very small. 
One can therefore use a threshold under which, matrix elements are set to zero in the reconstruction.

We first test the reconstruction of the matrices $\widehat{B}$\,, $\widehat{CD}$\,, $\widehat{CED}$ in Fig.\ref{fig1}\,. 
Here, we consider the weighted directed network shown in Fig.~\ref{fig1}(a) that has $N_f=9$ followers and $N_l=1$ leader whose coupling is randomly obtained as described in the caption of Fig.~\ref{fig1}\,. 
One observes that the coupling within the followers  {and their overall degree} given by $B$ are accurately inferred in Fig.~\ref{fig1}(b)\,,  {i.e. both the topology and the weights are reconstructed.}
In Fig.~\ref{fig1}(c)\,, the matrix $CD$ is well reconstructed despite its smaller elements compared to $B$\,. 
One clearly identifies two groups of weights: one being close to zero corresponding to vanishing matrix elements in $CD$\,; the other group being in the interval $[0.01,0.03]$ which corresponds to the non-vanishing elements of $CD$\,. 
Moving on to Fig.~\ref{fig1}(d)\,, the reconstruction of $CED$ seems less accurate than the two previous matrices. 
While many matrix elements are correctly inferred, some vanishing elements might be reconstructed as non-zero. 
This is due to the relatively small amplitude of the elements of $CED$ and the finite length of the time-series. 
We purposefully chose time-series that were not too long, i.e. $N_t=5\times 10^5$ to showcase that the theory also provides useful information when one is not in the asymptotic limit. 
In the Supplemental Material~\cite{SM}, we show the error is smaller when the length of the time-series is increased. 
Also, in principle, because of the approximation in Eq.~(\ref{eqmain})\,, one does not expect a perfect match between the estimated matrices and the actual ones. 

Then, using the reconstructed matrix $\widehat{B}$\,, we obtain $\widehat{C}$ and $\widehat{D}$ in Fig.~\ref{fig1}(e), (f)\,. 
Both couplings from followers to the leader and from the leader to the followers are well reconstructed {, i.e. the topology and the weights are recovered.}
Eventually, we use Eq.~(\ref{est1}) to obtain $\widehat{E} = -0.371$\,, while the actual value is $E = -0.435246$\,.
Note that the standard deviation on $\widehat{E}$ over $\mathcal{M}$ is $0.0635$\,.  
This allows us to estimate the internal parameter of the leader using Eq.~(\ref{est4})\,, which gives $\widehat{\alpha_{N_l}} = 0.1564$\,, while the actual value is $\alpha_{N_l} = 0.1$~\cite{note1}\,. 
It is important to remark that, in the numerical example we show here $|E| = 0.435246$\,, which is not close to zero as assumed in Eq.(\ref{eqmain})\,. 
Interestingly, even though the leader agent has a finite memory, the truncation used in Eq.(\ref{eqmain}) is still accurate enough to fairly reconstruct all the four blocks of $A$\,. 
{To further explore the validity of the approximation Eq.~(\ref{eqmain})\,, we vary the value of $E$ in Fig.~\ref{fig1_2}\,, and show the error between the reconstructed matrices and the actual ones. 
As expected, for both the matrix products (upper panel) and the blocks (lower panel), the error increases with $|E|$\,.}
Potential improvement could be achieved by truncating Eq.(\ref{eq3}) at a higher power of $E$\,.
{In the example shown here, the error in the reconstruction of $C$ and $D$ remains remarkably low, even for values of $E$ close to unity~\cite{SM}.}

\begin{figure}
    \centering
    \includegraphics[width=0.99\linewidth]{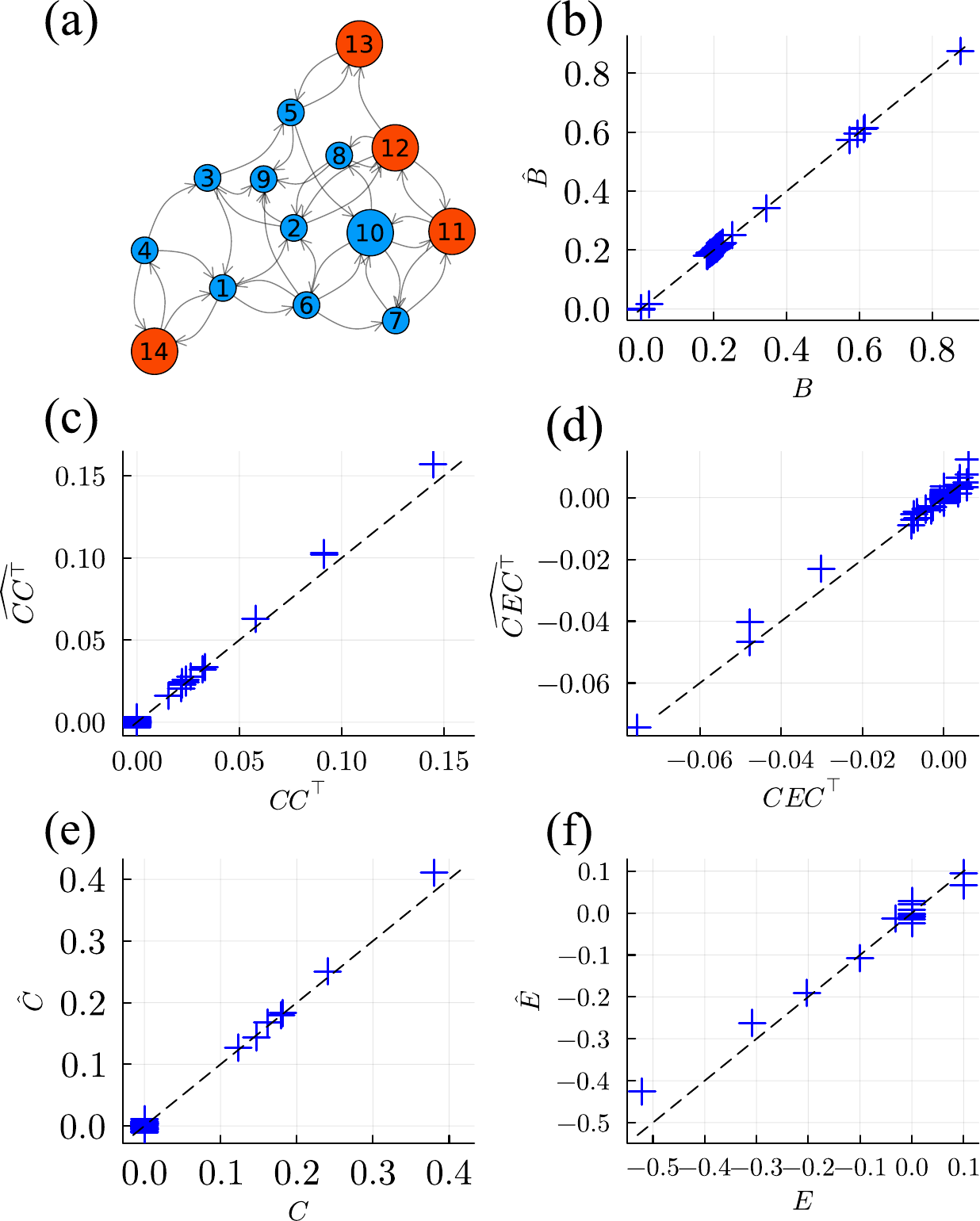}
    \caption{\textbf{Matrix reconstruction for multiple hidden leaders}. (a) Directed network of $14$ nodes with $N_f=10$ followers (blue) and $N_l=4$ hidden leader (red). 
    (b) Comparison between the actual coupling among the followers $B$ and the reconstructed one $\widehat{B}$\,. 
    (c) Comparison between the actual matrix $CC^\top$ and the reconstructed one $\widehat{CC^\top}$\,. 
    (d) Comparison between the actual matrix $CEC^\top$ and the reconstructed one $\widehat{CEC^\top}$\,. 
    The time-series for the reconstructions were obtained by simulating the dynamics Eqs.(\ref{eq1a}), (\ref{eq1b}) with the coupling given by the weighted network shown in panel (a). 
    The adjacency matrix of the network was obtained starting from a matrix of uniform random number between $0$ and $1$ and then keeping only the elements larger than $0.8$ and ignoring the diagonal. 
    The Laplacian matrix is then obtained from this adjacency matrix and normalized by the largest diagonal element. 
    The leader-follower coupling $C$\,, has been chosen to be symmetric, $D = C^\top$\,. 
    The internal parameters of the leaders are $(\alpha_{11},\alpha_{12},\alpha_{13}, \alpha_{14}) = (0.2,0.1,0.05,0.1)$\,. 
    The length of the time-series is $N_t = 1\times 10^6$\,, with vanishing initial conditions. 
     {The noise standard deviation is taken as 0.1 for all followers~\cite{note2}.}}
    \label{fig2}
\end{figure}

\section{Multiple unobserved leaders}
In general, when more than one leader is hidden, it is more complicated to fully reconstruct the matrix $A$\,. 
Indeed, even if one can still reconstruct the matrices $\widehat{B}$\,, $\widehat{CD}$\,, $\widehat{CED}$\,, without some additional assumption on the connectivity within the system, in general one cannot uniquely recover $C$\,, $D$ and $E$\,. 
Here, to lift that degeneracy, we will assume that the leaders are (i) symmetrically coupled to at least one follower each, i.e. $D = C^\top$\,; (ii) the leaders are not connected to the same followers. 
Note that $B$ does not have to be symmetric under these assumptions. 
Let us have a closer look at these three assumptions. 
Moreover, in order for Eq.~(\ref{eqmain}) to be valid, one then needs $|E_{ii}|=|\alpha_{N_f+i} - \kappa_i|\ll 1$ for $i=1,...,N_l$\,. 
Both assumption (ii) and (iii) together allow to unambiguously reconstruct $C$ and therefore also $E$ (see next section). 
Indeed, once the matrix $\widehat{CC^\top}$ has been obtained, one can reconstruct $C$ by identifying all the sets of non-vanishing columns of $\widehat{CC^\top}$ that are linearly dependent. 
Then, by picking only a single column for each set, one can reconstruct $C$ as~\cite{SM},
\begin{align}\label{eq6}
    \widehat{C}_{:,i}&=
      \frac{{\widehat{CC^\top}}_{:,j(i)}}{\left({\widehat{CC^\top}}_{j(i)j(i)}\right)^{1/2}}
\,,
\end{align}
for $i=1,...,N_l$\,, where ${\widehat{CC^\top}}_{:,j(i)}$ denotes the $j(i)$-th column of $\widehat{CC^\top}$\,, and $j(i)$ maps the column of $\widehat{C}$ to the selected columns of $\widehat{CC^\top}$. 
Doing so, one obtains an estimate of $C$ up to a permutation of its columns, which corresponds to a permutation of the indices of the hidden leaders (see below).

\begin{figure}
    \centering
    \includegraphics[width=0.99\linewidth]{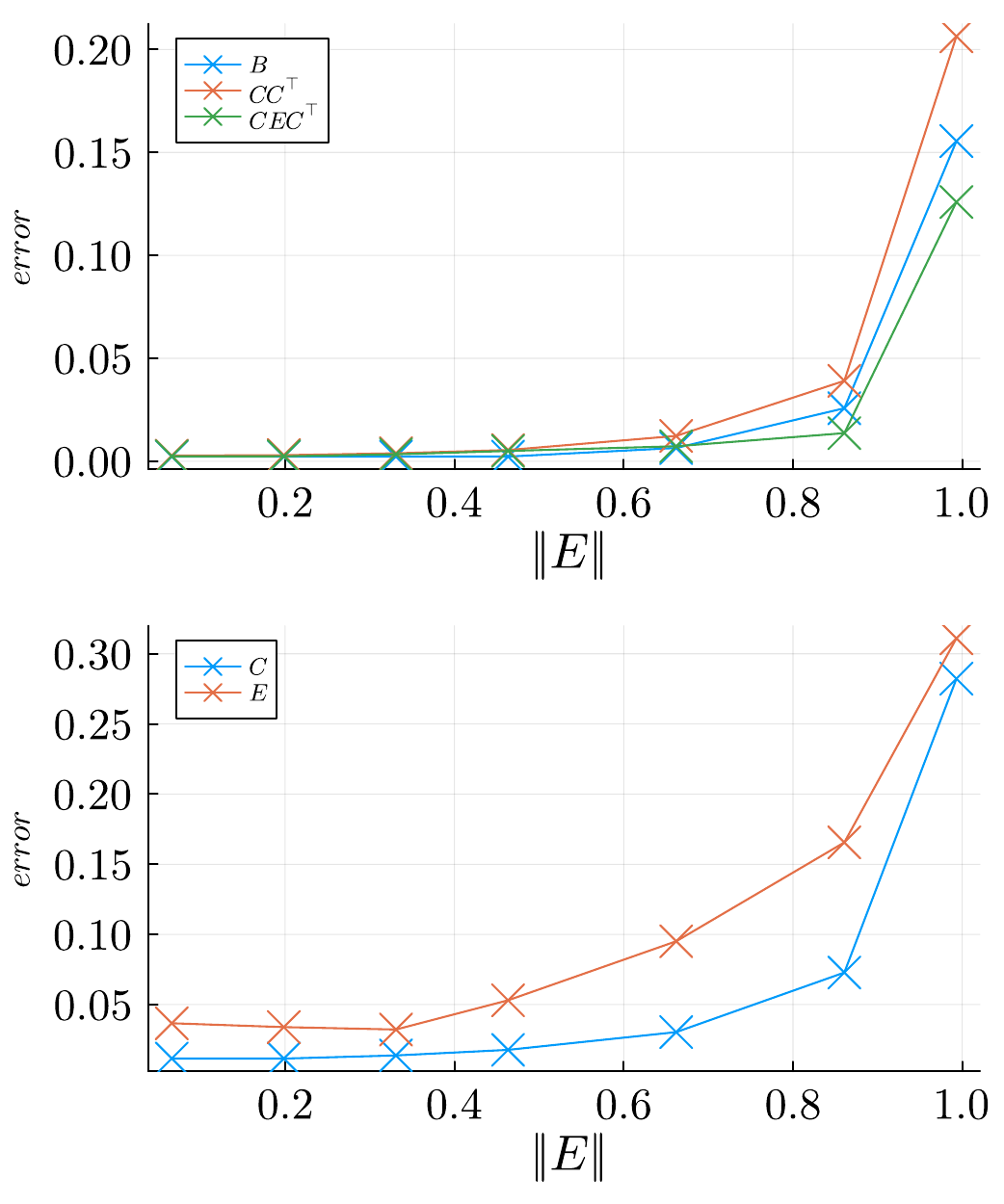}
    \caption{{\textbf{Error on matrix reconstruction for multiple hidden leaders}. Maximum of the absolute difference between the actual matrices and the reconstructed ones for different values of $E$\,. 
    The parameters of the system are the same as in Fig.~\ref{fig2}\,, but with $E$ that is multiplied by an overall factor that scales its Froebinius norm $\|E\|$\,. 
    The error can be compared to the absolute value of the non-vanishing elements of each matrix without the multiplicative factor over $E$\,: $|B_{ij}|=[0.023,0.877]$\,, $|[CC^\top]_{ij}|=[0.015,0.145]$\,, $|[CEC^\top]_{ij}|\in[0.001,0.075]$\,, $|C_{ij}|\in [0.12,0.38]$\,, $|E_{ij}|\in[0.032,0.521]$\,.    
    The length of the time-series is $N_t = 10^6$\,, with vanishing initial conditions. }}
    \label{fig2_2}
\end{figure}

Then, using $\widehat{C}$ and $\widehat{CEC^\top}$\,, one achieves the reconstruction of $E$ by solving the overdetermined system,
\begin{align}\label{eqml1}
    \widehat{C}\widehat{E}\widehat{C}^\top&= \widehat{CEC^\top}\,.
\end{align}
This can be done using the pseudo-inverse of $\widehat{C}$\,. 
Eventually, like we did in the single hidden leader case, one can obtain the internal leader dynamics from,
\begin{align}\label{eqml2}
    \widehat{\alpha_{N_f+i}} =  \sum_{j=1}^{N_f} {\widehat{C}_{ji}} + \sum_{j=1}^{N_l} {\widehat{E}_{ij}}\,,
\end{align}
for $i=1,...,N_l$\,. 
Now that we have estimators for all the block of the matrix $A$\,, let us test them on a numerical example. 
In Fig.~\ref{fig2}\,, we consider a network with $10$ follower agents and $4$ hidden leaders, where the leaders are symmetrically coupled to the followers [see Fig.~\ref{fig2}(a)]. 
Note that the followers are not symmetrically coupled with each others. 
As for the single hidden leader case, the matrices $\widehat{B}$\,, $\widehat{CC^\top}$\,, $\widehat{CEC^\top}$ are well reconstructed in Fig.~\ref{fig2}(b)-(d)\,. 
Then, using Eqs.~(\ref{eq6}), (\ref{eqml2})\,, the leader-follower coupling and the internal dynamics of the leaders are accurately inferred in Fig.~\ref{fig2}(e), (f)\,. 
The internal parameters of the leaders inferred as $(\widehat{\alpha_{11}},\widehat{\alpha_{12}},\widehat{\alpha_{13}},\widehat{\alpha_{14}}) = (0.32,0.09,0.07,0.141)$\,, while the actual parameters are $(\alpha_{11},\alpha_{12},\alpha_{13}, \alpha_{14}) = (0.2,0.1,0.05,0.1)$~\cite{note3}\,. 
Note that the spectral radius of $E$ is not much smaller than 1, meaning that the leaders do have some finite memory in the chosen example [$\rho(E)=0.561$]. 
{In Fig.~\ref{fig2_2}\,, we tune the spectral radius of $E$ and show the error between the reconstructed matrices and the actual ones. 
As observed in the single hidden leader case, the error increases with $\rho(E)$\,, which is expected from the approximation Eq.~(\ref{eqmain})~\cite{SM}.}

\section{Degeneracy in the reconstruction}
As discussed in the previous section, in order to reconstruct unambiguously the matrices $C$\,, $D$\,, $E$ from $B$\,, $CD$\,, $CED$\,, one needs to assume some properties about the coupling in the system. 
Indeed, let us discuss some transformation that would leave the martix products unchanged. 
For example, one can define alternative matrices $C'=CU$\,, $D'= U^{-1}D$\,, $E' = U^{-1}E U$ such that $C'D' = CD$\,, $C'E'D' = CED$\,.
From here, without further assumption on the structure of the coupling, one couldn't find whether the observed system is made of $C$\,, $D$\,, $E$\,, or $C'$\,, $D'$\,, $E'$\,.
\\
To lift this degeneracy, here we suggested the assumptions on the coupling (i), (ii), given in the previous section. 
Below, we further explain how the degeneracy is lifted. 
Note that, these assumptions are not the only ones that can lift the degeneracy in the reconstruction. 
Using assumption (i)\,, from the reconstructed matrices $\widehat{CC^\top}$\,, $\widehat{CE^nC^\top}$\,, $n>1$\,, one cannot tell whether there are obtained from $C$\,, $E$ or from $CU$\,, $U^\top E U$\,, where $U$ is a square $N_l\times N_l$ orthogonal matrix, i.e. $UU^\top = U^\top U = \mathbb{I}$\,. 
Let us assume that there is a degeneracy, namely, instead of reconstructing $C$\,, one reconstructs $\widehat{C}=CU$\,.
However, with assumption (ii), that hidden leaders are not connected to the same follower, each row of $C$ only has at most a single non-vanishing element. 
Therefore, one can write the reconstructed matrix as $\widehat{C}_{ij} = C_{ik}U_{kj}$ where the $i$-th row of $C$ has a non-vanishing elements at the $k$-th column, i.e. there is a connection between the $i$-th follower and $k$-th leader agents, which are respectively observed and hidden. 
Now, given Eq.~(\ref{eq6})\,, the matrix $\widehat{C}$ has the same structure (entries with non-vanishing elements) than the actual $C$ and $CU$\,, up to a permutation of the columns. 
In order to have such structure, one needs to have only a single non-vanishing elements for each row of $U$\,. 
Together with $U$ being orthogonal, this imposes the non-vanishing elements of $U$ to be unity, so that $U$ corresponds to a permutation of columns for $C$\,. 
This means that $C$ can be obtained by permuting the columns of $CU$\,. 
The degeneracy of the reconstruction is thus limited to the permutation of the indices of the hidden nodes which is irrelevant for the reconstruction of the dynamics. 
Therefore, with assumption (i) and (ii), one can recover the actual structure of the observed system.

\section{Conclusion}
We investigated a consensus algorithm where one has access to time-series measurement from a subset of agents, namely the followers. 
Solely based on these time-series, we proposed a method leveraging an autoregressive expansion of the observed dynamics that enables the reconstruction of the system's dynamics including both the observed and the hidden agents. 
Our method can be useful to identify the driver nodes in a complex network using only partial observations. 
In principle, the expansion can be truncated to the first two terms if the leader agents have a short memory. Numerically, we found that, even when the memory of the leaders is not short, the truncation provides an accurate reconstruction of the system's parameters. 
We anticipate that, by considering more terms in the approximation of the autoregressive expansion, one could obtain improved results.  
This is one of the future avenues to be explored, including also the extension of the multiple leaders case to the situation where the leader are not symmetrically coupled to the followers. The latter can be tackled using an SVD on the reconstructed matrix $\widehat{CD}$\,, but we found that the reconstruction is somewhat less accurate, and leave it to a future work. 
Also, here we assumed the knowledge of the Laplacian dynamics. One could investigate other types of additional information such as the network structure, without knowing the weights. 

\section{Acknowledgments}
We thank Andrey Lokhov, Marc Vuffray and Mateusz Wilinski for useful discussions.

%

\pagebreak
\widetext
\begin{center}
\textbf{\large Supplemental Material: Network Reconstruction in Consensus Algorithms with Hidden Agents}
\end{center}
\setcounter{equation}{0}
\setcounter{figure}{0}
\setcounter{table}{0}
\makeatletter
\renewcommand{\theequation}{S\arabic{equation}}
\renewcommand{\thefigure}{S\arabic{figure}}
\renewcommand{\bibnumfmt}[1]{[S#1]}
\renewcommand{\citenumfont}[1]{S#1}

\setcounter{figure}{0}
\renewcommand{\figurename}{Fig.}
\renewcommand{\thefigure}{S\arabic{figure}}

\subsection{Additional Figures}
\begin{figure}[h!]
    \centering
    \includegraphics[width=0.99\linewidth]{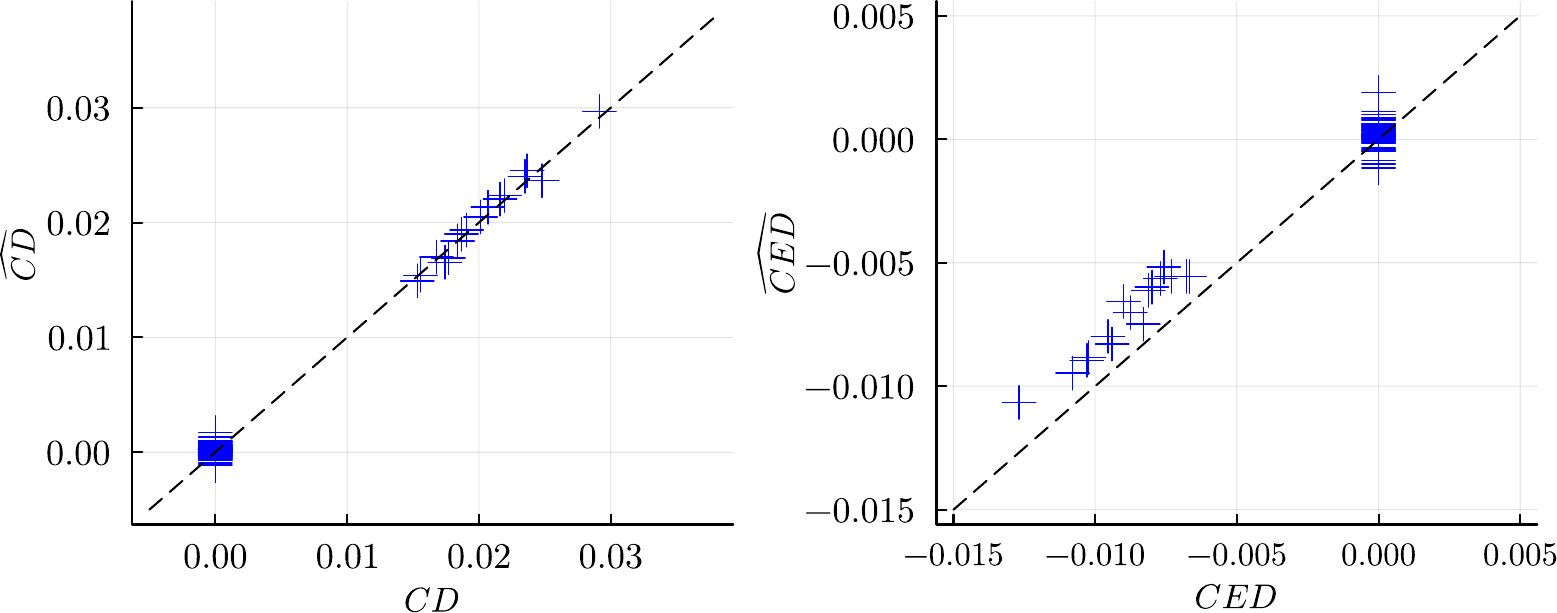}
    \caption{Same figure as panels (c) and (d) in  Fig.~1 in the main text, but with time-series of length $N_t = 5\times 10^6$\,. 
    Here, the distinction between the zero and non-zero elements in clear. 
    As expected from the truncation in Eq.~(5) in the main text, the weights converge to values that are not exactly the actual ones. }
    \label{fig:S1}
\end{figure}

\begin{figure}[h!]
    \centering
    \includegraphics[width=0.6\linewidth]{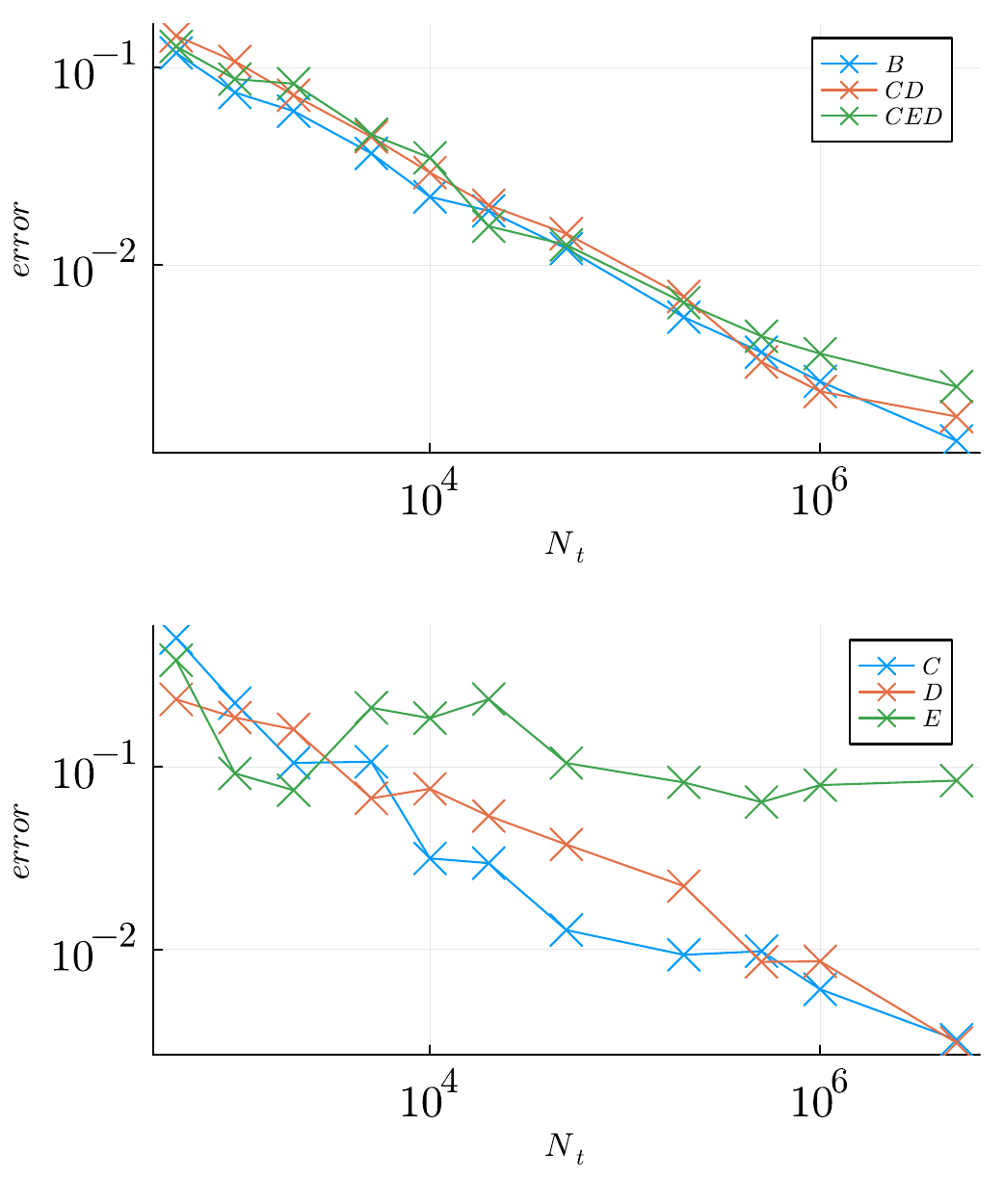}
    \caption{\textbf{Error as a function of $N_t$\,.} Maximum of the absolute difference between the reconstructed matrices and the actual ones. 
    The parameters of the system are the same as in Fig.~1 in the main text.}
    \label{fig:S2}
\end{figure}

Fig.~\ref{fig:S1} show the same system as in Fig.~1 in the main text, with longer time-series. 
One observes the improvement in the convergence of the estimates. 
However, as expected, in the right panel of Fig~\ref{fig:S1}\,, the elements of $\widehat{CED}$ do not converge to the actual elements of $CED$\,. This is due to the model  mismatch introduced by the approximation in Eq.~(5) in the main text.
This is further illustrated in Fig.~\ref{fig:S2} that shows the error as a function of the length of the time-series. 
While the error for the matrix products in the upper panel decreases as $N_t^{-1/2}$\,, the one observes that the error for the reconstructed $E$ seems to plateau. 
This is due to the model mismatch introduced by the approximation Eq.~(5) in the main text.
The same observations can be made in Fig.~\ref{fig:S2} that shows the evolution of the error with $N_t$ for the multiple hidden leaders case.

\begin{figure}[h!]
    \centering
    \includegraphics[width=0.6\linewidth]{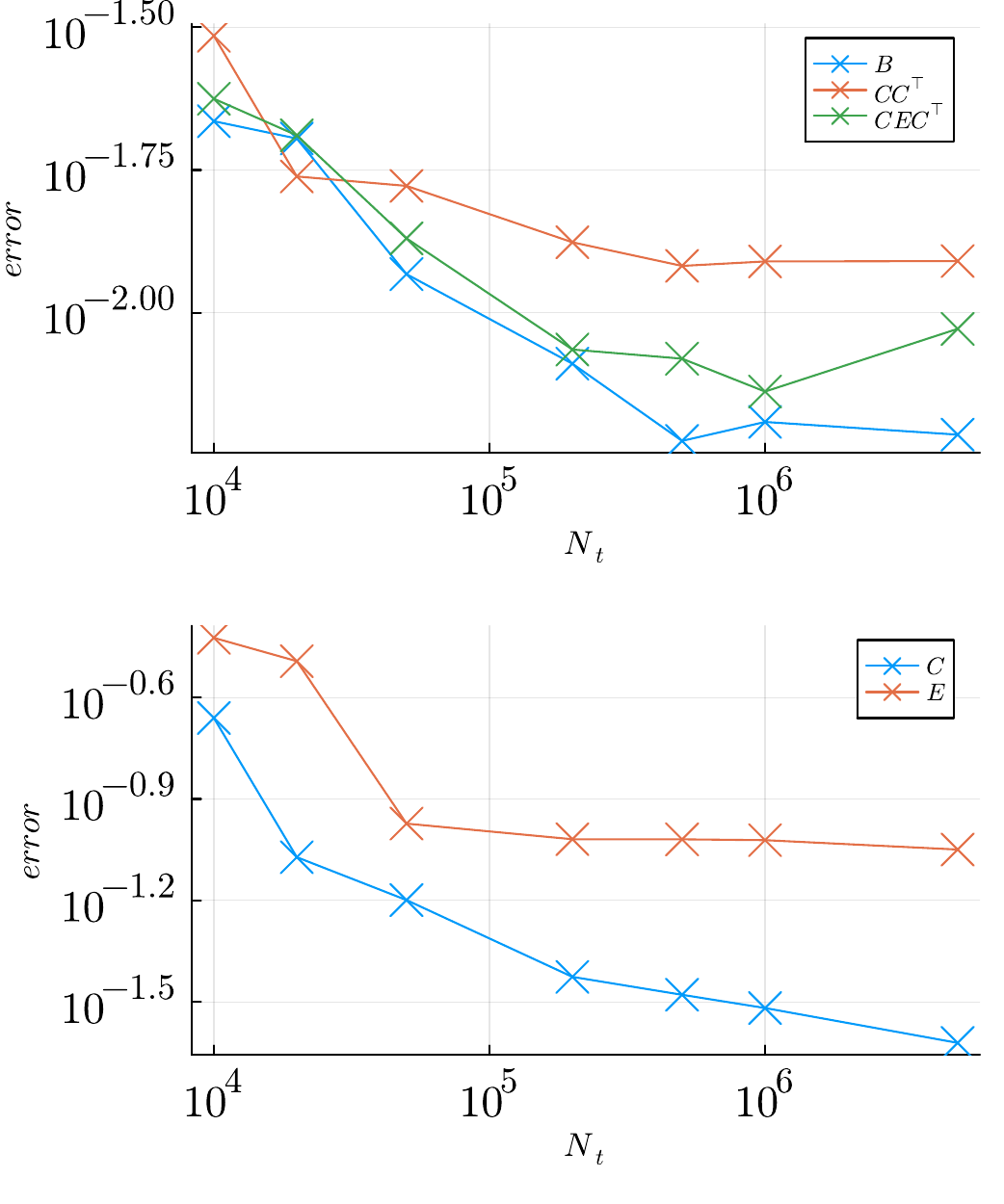}
    \caption{\textbf{Error as a function of $N_t$\,.} Maximum of the absolute difference between the reconstructed matrices and the actual ones. 
    The parameters of the system are the same as in Fig.~3 in the main text.}
    \label{fig:S3}
\end{figure}

\section{Multiple hidden leader reconstruction}
Here we provide more details about the reconstruction of the matrix $C$ for the multiple hidden leaders case. 
With the assumption given in the main text, the matrix $C$ has the following form,
\begin{align}
    C = \begin{pmatrix}
        a & 0 & 0 \\
        b & 0 & 0 \\
        0 & 0 & 0 \\
        0 & c & 0
    \end{pmatrix}
\end{align}
where, for illustrative purposes, we chose $C\in \mathbb{R}^{4\times3}$\,. 
Then, the matrix $CC^\top$ is given by,
\begin{align}
    CC^\top = \begin{pmatrix}
        a^2 & ab & 0 & 0 \\
        ab & b^2 & 0 & 0 \\
        0 & 0 & 0 & 0 \\
        0 & 0 & 0 & c^2
    \end{pmatrix}\,.
\end{align}
One remarks that the first two column of $CC^\top$ are linearly dependent. 
Therefore, by selecting the linearly independent columns of $CC^\top$ of dividing them by the diagonal element\,, one can recover the columns of $C$\,.



\begin{thebibliography}{31}
\bibitem [{not({\natexlab{a}})}]{note1}%
  \BibitemOpen
\bibfield  {journal} {  }\href@noop {} {\bibfield  {journal} {\bibinfo  {journal} {We run the inference for the same system and time-series length, but changing the noise sequence and found, over 20 realization, an average value of $\widehat{\alpha_{N_l}} = 0.199$ with standard deviation $0.015$\,.}\ } ({\natexlab{a}})}\BibitemShut {NoStop}%
\bibitem [{not({\natexlab{b}})}]{note2}%
  \BibitemOpen
  \href@noop {} {\bibfield  {journal} {\bibinfo  {journal} {The noise standard deviation does not need to be homogeneous over the followers. We tested the method with inhomogeneous standard deviations and found similar results.}\ } ({\natexlab{b}})}\BibitemShut {NoStop}%
\bibitem [{not({\natexlab{c}})}]{note3}%
  \BibitemOpen
  \href@noop {} {\bibfield  {journal} {\bibinfo  {journal} {We run the inference for the same system and time-series length, but changing the noise sequence and found, over 20 realization, average values of $(\widehat{\alpha_{11}},\widehat{\alpha_{12}},\widehat{\alpha_{13}},\widehat{\alpha_{14}}) = (0.3,0.09,0.06,0.12)$ with standard deviations respectively $(0.03, 0.06, 0.06, 0.03)$\,.}\ } ({\natexlab{c}})}\BibitemShut {NoStop}%
\end{thebibliography}

\end{document}